\documentstyle{article}

\begin{document}
\begin{center} {\Large {\bf On strangelets in magnetosphere}\footnote
{astro-ph/9410028; ifug-22/94}}\\[1cm]
Haret C. Rosu
\footnote{e-mail: rosu@ifug.ugto.mx} \\[2mm]
IFUG,
Apdo Postal E-143\\
Le\'on, Gto, M\'exico
\end{center}
$ $\\[1.5mm]
\begin{center}
{\bf Abstract}

A short discussion is provided on the hypothesis of strange quark
dust of low specific charge in the terrestrial magnetosphere. We suggest
by means of rough estimates the possible existence of a strangelet belt
in the magnetosphere.

\end{center}

$ $ \\[.7cm]
PACS: xx.xx.Let1let2, xx.xx.Let1let2\\[1cm]
\section{Introduction}

 The dark matter (DM) is a central paradigm of modern astrophysics with
 cosmological implications. It is based on the observed flat
 rotation curves of galaxies. Various exotic DM candidates
 have been proposed in the literature. Among them, baryonic DM was not much
 favored theoretically in the past few years. However recent possible
 detection by
 MACHO \cite{n2} and EROS \cite{n1} collaborations using the microlensing
 scanning technique
 of Large Magelanic Cloud (LMC) resettled the interest in baryonic DM \cite{j}.
 The statistics of
 microlensing events is rapidly increasing \cite{n3} and already an average
 mass of $\approx 0.1 M_{\odot}$ has been derived for the DM
 chunks. At the same time, the observed events
 might be ascribed to microlensing by stars within the LMC itself \cite{nn}.

My aim in this paper is to discuss strange quark matter (SQM),
a more exotic baryonic candidate for DM, in a space physics context,
namely within the magnetosphere, updating a hypothesis that I made 8
years ago \cite{r1}. The essential speculation is the possible existence of
a strangelet belt in the magnetosphere.

\section{Strange Quark Matter}

If we look strictly at the numerical value of the binding energy, the
most stable matter form known so far is nuclear matter ($\sim 15 MeV/N$)
and its clustered aspect- atomic nuclei ($\sim 8 MeV/N$). Quantum
Chromodynamics (QCD), the nonabelian generalization of Quantum
Electrodynamics (QED) has opened new horizons concerning metastable or
even stable matter states at high energy densities ($\geq 1 GeV/$fm$^{3}$).
As we all know, QCD asserts that the {\em atoms} of our world are
the so-called quarks, a kind of nonabelian electrons living only within
a limited volume of hadronic range ($\sim 1$ fm$^{3}$). We stress that we
do not know what quarks really are and all our speculations start by
considering them just ordinary fermions with a more complicated internal
symmetry. This fact might not be quite right. The chromophotons
(i.e., the gauge fields of QCD) are known as gluons and live also in the
compressed (perturbative) vacuum inside hadrons. This confinement property
of QCD is one of the two most important features of this theory. It is
not at all well understood at the present time. The other fundamental
property, asymptotic freedom, enables someone to apply perturbative
methods at very high energies. The arguments used to give reasons for
confinement are topological ones (e.g., the chromoelectric Meissner effect),
but in our opinion these are only analogies. If the confinement property
is questioned, considering it a function of some parameters (e.g., the
energy) we enter the wonderful world of all kinds of quark matter forms,
a great challenge for detailed studies. One of the most promising candidate
in this scheme is precisely SQM, possessing a surplus of
s quarks (q = -1/3 e) over the ordinary quark matter. It was Witten
\cite{w}, back in
1984, who claimed that the s quark surplus comes from assuming the Fermi
momentum of the quark matter to be about 300 MeV, more than the mass of
that quark ($\sim 200 MeV$) and thus allowing a part of the up and down
quarks to become spontaneously strange quarks. In this way the Fermi
momentum is lowered and so is the energy. A massless quark flavor taken
as a Fermi gas exerts a pressure $p_{F}^{4}/4\pi ^{2}$, where $p_{F}$ is
the Fermi momentum. For quark matter without strangeness, charge neutrality
implies a chemical potential $2^{1/3} \mu$ for the d quarks, where $\mu$ is
the chemical potential for the u quarks. Thus the total pressure would be
$(1+2^{4/3}) \mu ^{4}/4\pi ^{2}$. In the case of 3 flavors and zero mass for
s quarks, the same pressure is exerted at a lower Fermi momentum,
$\mu _{3}=[1/3(1+2^{4/3})]^{1/4} \mu$. When we compare this with the Fermi
momentum of a two flavor (u,d) quark matter, $\mu _{2}=[1/3(1+2^{4/3})]\mu$,
we get a figure smaller than unity, $\mu _{3}/\mu _{2}=0.89$. One can see
that a parameter window opens up for the possible existence of three
flavored quark matter. To go to more flavors is already not allowed, since
the subsequent c,b,t quarks are too heavy and only unstable matter forms
can be generated. The above Witten's estimates on the existence of SQM
have been taken seriously by various authors. De Rujula \cite{r}
and De Rujula and Glashow \cite{rg} speculated that forms of SQM may populate
the huge nuclear desert of 54 orders of magnitude in mass spectrum
(and dimensions) lieing between atomic nuclei and neutron stars.
Alcock and Farhi \cite{af} have shown that only lumps with baryon number
larger than
$10^{52}$ (planetary masses) could survive with decreasing temperature of
the Universe, close to the average mass of DM chunks derived by microlensing.
 Moreover, the astrophysical origin of SQM was investigated,
for example Cyg X-3 can be an astrophysical source \cite{wisc}.
Strangeness affecting the inner core of neutron stars is a well-known
conjecture \cite{bp}. Of course, one should imagine physical processes
by which SQM is reaching the crust of neutron stars, and even raised to
the pulsar magnetosphere giving signals to our antennas. An interesting
argument is that SQM forms from atomic up to meteoritical dimensions might
propagate in the astrophysical space over large distances, only slightly
affected by the galactic magnetic field, due to its quasi-neutrality.
So, even astrophysical SQM can be found in Solar System and in the very
neighborhood of our planet. We recall that in 1979, Bjorken and
McLerran \cite{bl} studied the hypothesis that Centauro cosmic ray events
are initiated by explosive quark matter, and in the same year, Chin and
Kerman \cite{ck} made an analysis of a class of strange quark droplets
using the MIT bag model of hadrons.

Witten \cite{w} gave the following estimate for the free SQM in our
galaxy. Since the neutron stars are about 1 \% of the star population
of a galaxy, this means some $10^{9}-10^{10}$ neutron stars in our
Milky Way. Free QSM may be produced astrophysically in the collisions
of neutron stars (very rare events) and in the gravitational disintegration
of binary neutron star systems ($10^{6}M_{\odot}$ in our galaxy). With
$10^{-1} M_{\odot}$ ejected from such binary systems, there might be
$10^{5}M_{\odot}$ of free SQM in our galaxy.

Farhi and Jaffe \cite{fj} provided the most detailed analysis of the
properties of SQM. They used a Fermi gas model with O($\alpha _{c}$)
corrections and a bag model in the region of low baryon number (near the
nuclear limit). The investigated baryon number spectrum ranged from
$10^{2}$ up to $10^{7}$ and the main parameters were $m_{s}$, $\alpha _{c}$,
and B, the strange quark mass, the QCD coupling constant, and the vacuum
energy density shift, respectively. Beginning with the PANIC talk of
De Rujula \cite{r} the notion
of nuclearites, i.e., strange quark balls, started circulating.

\section{Nuclearites in the magnetosphere}

In his 1984 PANIC talk, De Rujula \cite{r} gave some useful estimates
concerning the nuclearites. Rujula supposed that nuclearites are major
components of the local DM the Earth is passing through.
They have to interact with
matter by atomic collisions and there is energy loss. From this loss, the
range of nuclearites as a function of their mass can be obtained. De Rujula
found that nuclearites smaller than $4 10^{-14}$ grams cannot penetrate
the atmosphere, since they have very small ranges, and therefore at the
see level one can see them only indirectly through their secondaries. For
a direct detection one may think of the magnetospheric environment.
Before discussing this, let us recall that nuclearites have been classified
in two types \cite{fj}, \cite{b}. Those ones with
$A\leq 10^{6}$ amu = $10^{-18}$ grams are much akin to atoms, with clouds of
electrons (as many as several hundreds) around, neutralizing the charge,
which because of strangeness is only $Z=6 A^{1/3}$ instead of the usual
$Z\sim A/2$ for atomic nuclei. This type of nuclearites going down to the
nuclear limit ($10^{-22}$ grams) are called strangelets. As an example,
a strangelet of $10^{-18}$ grams has 600 electrons around and is 100 fm
in size. The larger nuclearites are not very different from metalic
ordinary matter, with the electrons inside them as a degenerate Fermi gas.
One can consider even meteoritic scales as long as nothing is known on
their mass distribution. In the following we shall concentrate on
strangelets as fitting better a magnetospheric context. As we said
strangelets can travel easily over large astrophysical distances. However
strong magnetic fields of magnetospheric type can capture them, in view of
the ionizing processes due to the particles of the radiation belts.
The nuclearite motions in the magnetosphere can be studied as a simple
application of the St\"ormer theory of the motion of charged
particles in a dipole magnetic field. Here we sketch a short resume following
the excellent review of Vallarta \cite{v}.

Take $\rho$, $\lambda$, and $\phi$ as spherical coordinates (radial distance,
latitude, and longitude). The origin is chosen in the center of the Earth
magnetic dipole. Positive $\lambda$ means northern hemisphere, and positive
$\phi$ is westwords. A very useful change of variables was discovered by
St\"ormer, through which all the physical quantities under consideration
lose their dimensions: $r=\rho/l_{S}$ and $ds=v/l_{S}dt$, where $l_{S}$ is the
St\"ormer parameter given by $l_{S}=\sqrt{Mq/mv}$, where $M$ is the magnetic
dipole moment, $q/m$ is the specific charge of the particle, and $v$, its
velocity.

In adimensional variables, magnetospheric movements of particles are written
down as
$$r^{''}-r\lambda ^{'2}-r\phi ^{'2}\cos ^{2} \lambda =
 -\frac{\cos ^{2} \lambda}{r^{2}}\phi ^{'}   \eqno(1a)$$
 $$r\lambda ^{''}+2r^{'}\lambda ^{'} +r\phi ^{'2}\sin \lambda \cos \lambda =
 -\frac{2 \sin \lambda \cos \lambda}{r^{2}}\phi ^{'} \eqno(1b)$$
$$\frac{1}{r\cos \lambda}(r^{2} \phi  ^{'}\cos ^{2} \lambda)^{'}=
\frac{2\sin \lambda}{r^{2}}\lambda ^{'}+
\frac{\cos \lambda}{r^{3}} r^{'}   \eqno(1c)                 $$
Derivatives are taken with respect to $s$, the adimensional trajectory line
element. To this system of equations, energy conservation must be added
$$r^{'2}+r^{2}\lambda ^{'2} + (r^{2} \cos ^{2} \lambda) \phi ^{'2} =1
  \eqno(2)  $$

An immediate integration of Eq. (1c) gives
$$r^{2}\phi ^{'}\cos^{2} \lambda + \frac{1}{r}\cos ^{2}\lambda =const=
2\gamma _{1}   \eqno(3)  $$
where the constant of integration is another well known parameter in the
St\"ormer problem, which physically has the meaning of the axial projection
of the angular momentum on the dipole axis. Finally, one can easily obtain
the celebrated equation
$$ \pm \sin \theta = \frac{2\gamma _{1}}{r\cos \lambda} -
\frac{\cos \lambda}{r^{2}}    \eqno(4)  $$
where $\theta$ is the angle between the trajectory direction and the meridian
plane, and the $\pm$ sign is the charge sign. The last equation gives us
allowed and forbidden regions in the meridional plane corresponding to the
simple condition $|\sin \theta |\leq 1$.

As is known an experimental confirmation of the old St\"ormer theory (1907)
came only in 1958 after the first extraatmospheric flights, giving evidence
on the existence of radiation belts \cite{vA}. In the physics of belts the
McIlwain parameter $L=R_{E}/\cos ^{2} \lambda$ is used. The inner belt
(protons and a small number of heavy ions) lies at $1.6 \leq L_{1}\leq 2.1$,
while the second belt (mainly electrons) is to be found at
$3\leq L_{2} \leq 5$.

The strangelets, if any, should have a stationary behavior close to that of
heavy ions,
only that they should be found at even lower $L$ parameter than $L_{1}$.
Since their specific charge can be quite low perhaps they lase along magnetic
field lines in mirror movements similar to the ordinary belts. The
astrophysical strangelets can reach the lower $Ls$ by radial diffusion from
higher $L$ shells.

One may ask about the SQM fluxes. Using Witten's estimates, I found
$\sim 100$ grams of astrophysical SQM in the Solar System and for a mass
distribution with a strong peak at $10^{-21}$ grams, this would mean only
one Avogadro number of astrophysical SQM in the Solar System. At the Earth
orbit the corresponding figure would be $10^{-11}$ strgls/cm$^{2}$ s, quite
disappointing as expected. The DM hypothesis ($10^{-17}$ grams/cm$^{2}$s,
on the other hand, for the same mass distribution, allows for figures
comparable with the experimental ones for ordinary baryonic matter, i.e.,
$10^{4}$ protons/cm$^{2}$s, and $10^{1}$ alpha particles in the same units,
within the inner belt.

One can also be interested in an optimal value for the $L$ parameter (as a
function of energy) at which strangelet intensity reaches a maximum. The
estimation of this critical $L$ is known in the case of protons and heavy
ions \cite{pan}; for protons, $ L_{cr,p}=6.75 E^{-1/4}(MeV)$, and for heavy
ions, $L_{cr,hi}\sim Z^{1/2}A^{-1/4}E^{-1/4} (MeV)$. The same type of
relationship might work for strangelets. I give the following estimate,
$L_{cr, strgl}\sim 0.48Z^{3/2}A^{-1/4}E^{-1/4} (MeV)$.

\section{Conclusions}

A heuristic discussion of a possible belt in the magnetosphere made of
strangelets of low specific charge, similar to the more common radiation
belts has been given in this paper. Such a strangelet belt may be due to
the local DM density, and a very small component to astrophysical sources.
If taken seriously (indeed, the old result of Alcock and Farhi \cite{af}
is against our framework, as we are left with the astrophysical component
only), more detailed investigations will be required. One
can also go with this hypothesis to other accessible planetary magnetospheres,
e.g., Jupiter one.



\begin{thebibliography} {99}
\bibitem{n2} C. Alcock et al., Nature {\bf 365}, 621 (1993)
\bibitem{n1} E. Aubourg et al., Nature {\bf 365}, 623 (1993)
\bibitem{j} F. De Paolis et al., astro-ph/9410016.
\bibitem{n3} A. Udalski et al., Acta Astron. {\bf 43}, 289 (1993)
\bibitem{nn} K.C. Sahu, Nature {\bf 370}, 275 (1994)
\bibitem{r1} H. Rosu, in {\em Topics in Astronomy, Astrophysics, and
Space Sciences}, vol. ll, pp. 56-62 (CIP Press, Bucharest, 1986);
preprint A-13/March 1986, IFA-Magurele.
\bibitem{w} E. Witten, Phys. Rev. D {\bf 30}, 272 (1984)
\bibitem{r} A. De Rujula, Nucl. Phys. A {\bf 434}, 605 (1985)
\bibitem{rg} A. De Rujula and S.L. Glashow, Nature {\bf 312}, 734 (1984)
\bibitem{af} C. Alcock and E. Farhi, preprint CTP 1256 (1985)
\bibitem{wisc} V. Barger et al., Eds, {\em Cyg X-3}, Wisconsin report (1985)
\bibitem{bp} G. Baym and L.D. Pethick, Ann. Rev. Nucl. Sci. {\bf 25}, 27
(1975)
\bibitem{bl} J.D. Bjorken and L.D. McLerran, Phys. Rev. D {\bf 20}, 2353 (1979)
\bibitem{ck} S.A. Chin and A.K. Kerman, Phys. Rev. Lett. {\bf 43}, 1292 (1979)
\bibitem{fj} E. Farhi and R.L. Jaffe, Phys. Rev. D {\bf 30}, 2379 (1984)
\bibitem{b} L.J. Boya et al., Phys. Rev. A {\bf 32}, 1299 (1985)
\bibitem{v} M. Vallarta, Hand. d. Physik, vol. XLVI/1, pp. 88-129 (1961)
\bibitem{vA} J. van Allen and L.A. Frank, Nature {\bf 184}, 219 (1959)
\bibitem{pan} M.I. Panasiuk, Kosmicheskie Issledovaniya {\bf 18}, 83 (1980)
\end{thebibliography}
\end{document}